# Polarized photocathodes make the grade[*]

Jym Clendenin and Takashi Maruyama

*Stanford Linear Accelerator Center, 2575 Sand Hill Rd., Menlo Park, CA 94025*

**Abstract**

Future linear colliders will require high levels of performance from their electron sources. A group at SLAC has recently tested a structure that substantially exceeds current collider polarized electron source pulse-profile requirements.

---

[*]Work supported by Department of Energy contract DE–AC03–76SF00515.

# Polarized photocathodes make the grade

**Author**: *Jym Clendenin and Takashi Maruyama, SLAC*

Future linear colliders will require high levels of performance from their electron sources. A group at SLAC has recently tested a structure that substantially exceeds current collider polarized electron source pulse-profile requirements.

A polarized electron source for future electron-positron linear colliders must have at least 80% polarization and high operational efficiency. The source must also meet the collider pulse profile requirements (charge, charge distribution and repetition rate). Recent results from the Stanford Linear Accelerator Center (SLAC) have demonstrated for the first time that the profile required for a high-polarization beam can be produced.

Since the introduction in 1978 of semiconductor photocathodes for accelerator applications, there has been significant progress in improving their performance. Currently, all polarized electron sources used for accelerated beams share several common design features – the use of negative-electron-affinity semiconductor photocathodes excited by a laser matched to the semiconductor band gap, the cathode biased at between –60 and –120 kV DC, and a carefully designed vacuum system. While the earliest polarizations achieved were much less than 50%, several accelerator centers, including Jefferson Lab, MIT Bates and SLAC in the US, along with Bonn and Mainz in Germany, now routinely achieve polarizations of around 80%. Source efficiencies have shown similar dramatic improvement. The Stanford Linear Collider (SLC) achieved more than 95% overall availability of the polarized beam across nearly seven years of continuous operation. These achievements clearly point to the viability of polarized beams for future colliders.

Peak currents of up to 10 A were routinely produced in 1991 in the SLC Gun Test Laboratory by using the 2 ns pulse from a doubled Nd:YAG laser to fully illuminate the 14 mm diameter active area of a GaAs photocathode. However, when the photocathode gun was moved to the linac injector, where a high-peak-energy pulsed laser was available that could be tuned to the band gap energy as required for high polarization, the current extracted from the cathode was found to saturate at much less than 5 A unless the cathode quantum efficiency (QE) was very high.

The SLC required a source pulse structure of about 8 nC in each of two bunches separated by some 60 ns at a maximum repetition rate of 120 Hz. These requirements were met by doubling the cathode area and by using a vacuum load-lock to insure a high QE when installing newly activated cathodes. In contrast, designs for the Next Linear Collider and Japan Linear Collider, being pursued by SLAC and the KEK laboratory in Japan, call for a train of 190 microbunches separated by 1.4 ns, with each bunch having a 2.2 nC charge at the source, for a total of 420 nC for the 266 ns macropulse. This is about 25 times the SLC maximum charge. Both the macrobunch and microbunch current requirements for CERN's CLIC concept are somewhat higher, while the 337 ns spacing between microbunches insures that charge will not be a limitation for the TESLA collider being spearheaded by Germany's DESY laboratory.

The limitation in peak current density, which has become known as the surface charge limit (SCL), proved difficult to overcome. Simply designing a semiconductor structure with a high quantum yield was not a solution because the polarization tended to vary inversely with the maximum yield.

**Gradient doping**

As early as 1992, a group from KEK, Nagoya University and the NEC company designed a GaAs-AlGaAs superlattice with a thin, very-highly-doped surface layer and a lower density doping in the remaining active layer – a technique called gradient doping. The very high doping aids the recombination of the minority carriers trapped at the surface that increase the surface barrier in proportion to the arrival rate of photoexcited conduction band (CB) electrons. Because CB electrons depolarize as they diffuse to the surface of heavily doped materials, the highly doped layer must be very thin, typically no more than a few nanometers. When tested at Nagoya and SLAC, this cathode design yielded promising results in which a charge of 32 nC in a 2 ns bunch was extracted from a 14 mm diameter area, limited by the space charge limit of the 120 kV gun at SLAC.

In 1998 a group from KEK, Nagoya, NEC and Osaka University applied the gradient-doping technique to a strained InGaAs-AlGaAs superlattice structure. They retained 73% polarization while demonstrating the absence of the SCL in a string of four 12 ns microbunches, spaced 25 ns apart, up to the 20 nC space charge limit of the 70 kV gun. In a more recent experiment using a gradient-doped GaAs-GaAsP superlattice, they extracted 1 nC for each of a pair of 0.7 ns bunches separated by 2.8 ns without any sign of the SCL, before reaching the space charge limit of the 50 kV gun. The polarization and QE were 80 and 0.4% respectively. Other groups, notably at Stanford University, St. Petersburg Technical University and the Institute for Semiconductor Physics at Novosibirsk, have also made significant contributions to solving the SCL problem.

A group at SLAC has recently applied the gradient-doping technique to a single strained-layer GaAs-GaAsP structure with results that substantially exceed current collider requirements. These results both complement and extend the 1998 Japanese results. The highly doped surface layer was estimated to be 10 nm thick. To compensate for an increase in the bandgap that resulted from the increased dopant concentration, 5% phosphorus (P) was added to the active layer and the percentage of P in the base layer was increased to maintain the desired degree of lattice strain at the interface. Adding P in the active layer shifts the bandgap by about 50 meV towards the blue, reaching 1.55 eV (800 nm). In combination with the reduction of the surface barrier, this ensured a high QE of about 0.3% at the polarization peak. This is similar to the QE of the standard SLC strained GaAs-GaAsP cathodes.

Two laser systems were used to determine the peak charge. A flashlamp-pumped Ti:sapphire (flash-Ti) system provided flat pulses up to several hundred nanoseconds long with a maximum energy of about 2 μJ/ns. In addition, up to 20 μJ in a 4 ns pulse was available from a Q-switched, cavity-dumped, YAG-pumped Ti:sapphire (YAG-Ti) laser. With the flash-Ti alone, the charge increased linearly with the laser energy up to the maximum available laser energy. Because of the finite relaxation time of the SCL, a flat pulse is a much more stringent test of the SCL than if it contained a microstructure. The peak charge per unit time (see graph) is only slightly lower than the NLC requirement for each microbunch when assuming a 0.5 ns full bunch-width. By extending the laser pulse to 370 ns, a charge of 1280 nC was extracted, far exceeding the NLC macropulse requirement.

To determine if the peak charge required for a microbunch would be charge-limited, the YAG-Ti laser pulse was superimposed on the flash-Ti pulse. The resulting charge increment was consistent with the charge obtained using the YAG-Ti alone. The charge increment was independent of the relative temporal positions of the two laser pulses indicating that the massive total charge of an NLC, JLC or CLIC macropulse will not inhibit the peak charge required for each microbunch. The maximum charge produced by the YAG-Ti alone was 37 nC, which is more than 15 times the NLC requirement for a single microbunch.

To increase the charge density the laser spot on the cathode was reduced to 14 mm, below which the bunch is space-charge-limited for the maximum laser energy. Again, the charge increased linearly with the laser energy. The linearity remained when the quantum yield was allowed to decrease although, of course, the maximum charge also decreased. Thus it is clear that if sufficient laser energy is available, the linearity of the charge increase will be maintained for total charge and peak charge per unit time when using the new SLAC cathode design and will exceed NLC, JLC, and CLIC requirements.

The new SLAC cathode was used in the polarized source for a recent high-energy physics experiment requiring 80 nC at the source in a 300 ns pulse. The improved charge performance provided the headroom necessary for temporal shaping of the laser pulse to allow adequate compensation for energy beam loading effects in the 50 GeV linac. The polarization measured at 50 GeV confirmed the greater than 80% polarization measured in the source development laboratory at 120 keV.

The international effort to improve polarized photocathodes will continue. For instance, tests for the surface charge limit at the very high current densities required by low-emittance guns have yet to be performed. On a broader front, the superlattice structure — in part because of the large number of parameters that the designer can vary — appears to be the best candidate for achieving a significantly higher polarization while maintaining a QE above 0.1%.

**Further reading**

T. Abe *et al.* (American Linear Collider Working Group) 2001 Linear Collider Physics Resource Book for Snowmass 2001 (hep-ex/0106055-8 at http://www.arxiv.org/ ).

T. Maruyama *et al*. 2002 A very-high-charge, high-polarization gradient-doped strained GaAs photocathode (SLAC-PUB-9133) accepted for publication in *Nucl. Instrum. Meth. A*.

K. Togawa *et al.* 1998 Surface charge limit in NEA superlattice photocathodes of polarized electron source *Nucl. Instrum. Meth*. A414 431.

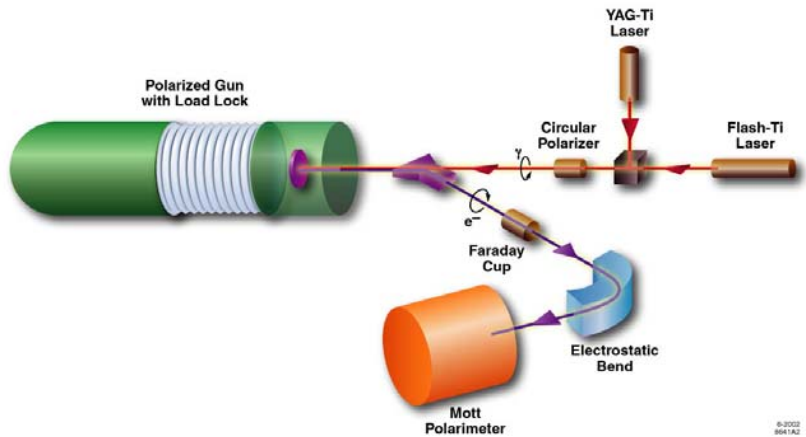

The configuration for SLAC's photocathode experiment.

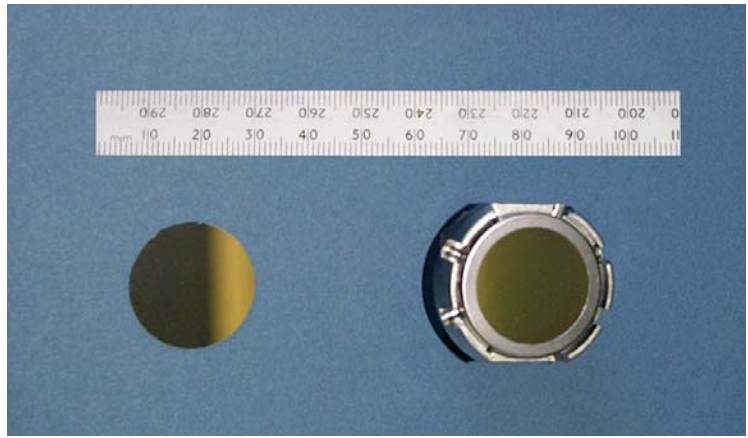

A photocathode crystal before (left) and after mounting in the crystal holder of the SLAC polarized electron source.

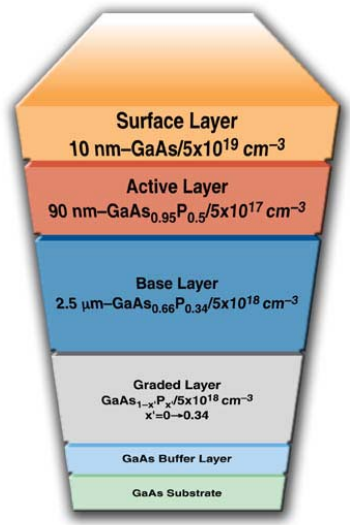

The high-gradient GaAs-GaAsP cathode structure, thickness and dopant density that was used for SLAC's polarized photocathode experiment.

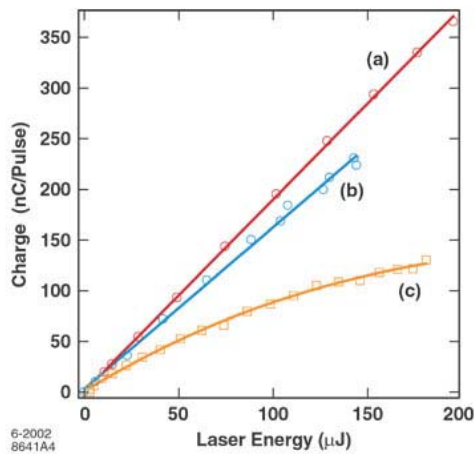

The charge in the electron bunch measured at the electron source as a function of laser energy using a 100 ns pulse with no microstructure: (a) QE of 0.31% and fully illuminated cathode diameter of 20 mm; (b) 0.25% and 14 mm; (c) SLC cathode shown for comparison: 300 ns pulse with QE of about 0.2% (at 10 nC) and 20 mm diameter.